\documentclass[aps,prl,twocolumn,twoside,showpacs]{revtex4}
\usepackage{amsfonts}

\begin{document}

\title{
 Massive Gauge Bosons in Yang-Mills Theory without
Higgs Mechanism }
\author{ Xin-Bing Huang}
\email[]{huangxb@pku.edu.cn} \affiliation{ Department of Physics,
Peking University,   100871 Beijing, the People's Republic of
China}
\date{September 28, 2004}

\begin{abstract}

A new mechanism giving the massive gauge bosons in Yang-Mills
theory is proposed in this letter. The masses of intermediate
vector bosons can be automatically given without introducing Higgs
scalar boson. Furthermore the relation between the masses of
bosons and the fermion mass matrix is obtained. It is discussed
that the theory should be renormalizable.
\end{abstract}

\pacs{11.15.Ex, 12.10.Kt, 12.60.Cn}

\maketitle

Half a century ago, Yang and Mills constructed the gauge field
theory of non-Abelian group, which has become the most fundamental
content in modern physics. Upon the principle that physical laws
should be covariant under the local isospin rotation they proposed
the SU(2) Yang-Mills theory~\cite{yan54}. But they could not
obtain the massive gauge bosons then. About 10 years later, An
ingenious trick called the Higgs mechanism was independently
invented by Higgs and Englert and Brout~\cite{hig64}, who
introduced a scalar field and the spontaneous symmetry broken
mechanism of vacuum by fixing a vacuum expectation value of the
scalar field and make the intermediate vector bosons obtain
masses. Based on the Yang-Mills fields and the Higgs mechanism,
Glashow, Salam and Weinberg {\em etc.} proposed a renormalizable
theory unifying the weak and electromagnetic interactions, namely
$SU(2)_{L}\times U(1)_{Y}$ gauge theory~\cite{gsw60}. In this
theory, the neutrinos are assumed to be massless and other
fermions can acquire the masses through the Yukawa couplings with
the Higgs boson.

Although the $SU(2)_{L}\times U(1)_{Y}$ electroweak theory has
predicted the masses of intermediate vector bosons, which are
confirmed by experiments, there are still several unconfirmed
predictions or conflicting phenomena in it. e.g. firstly,
experimenters do not find any hints of the Higgs boson till now;
secondly, several recent experiments imply that the neutrinos
should be massive and be mixed~\cite{xin04}. The neutrino
experiments demonstrate that people should modify the
$SU(2)_{L}\times U(1)_{Y}$ electroweak theory to be consistent
with more experimental phenomena. In this letter we propose a
mechanism to give the massive gauge bosons in Yang-Mills theory
without introducing a scalar boson. The equations connecting the
masses of intermediate vector bosons and the fermion mass matrix
are also obtained.

We demonstrate our mechanism under the primary framework of
Yang-mills theory in this letter. Let's consider a quantum field
system in which the fermions $\psi_{1}(x)$ and $\psi_{2}(x)$ form
an isospin doublet as follows
\begin{eqnarray}
\label{y0002} \psi= \left(
\begin{array}{c}
\psi_{1}(x)
\\
\psi_{2}(x)
\end{array}
\right)~.
\end{eqnarray}
While ignoring the helicity of fermions, the largest inner gauge
symmetry group in this system is obviously $SU(2)\times U(1)$
group. The complete Lagrangian of this system without mass terms
reads
\begin{eqnarray}\nonumber
 {\cal L}&=&i{\bar
{\psi}}\gamma^{\mu}(\partial_{\mu}-i g {\bf T}\cdot {\bf
A}_{\mu})\psi+g^{\prime}{\bar{\psi}}\gamma^{\mu}A^{0}_{\mu}\psi
\\
\label{y0001+} &-&\frac{1}{4}~{\bf F}^{\prime}_{\mu\nu}\cdot{\bf
F}^{\prime\mu\nu}-~\frac{1}{4}~E^{\prime}_{\mu\nu}E^{\prime\mu\nu}~.
\end{eqnarray}
Where $g$, $g^{\prime}$ are the coupling constants of $SU(2)$
gauge field and $U(1)$ gauge field respectively, and the dot
``$\cdot$" denotes a scalar product in the isospace. In this case,
${\bf T}\cdot {\bf A}_{\mu}$ means
$T_{1}A^{1}_{\mu}+T_{2}A^{2}_{\mu}+T_{3}A^{3}_{\mu}$, where
$T_{i}$ are the generators of $SU(2)$ group. In Yang-Mills theory,
${\bf T}\cdot {\bf A}_{\mu}$ is called the $SU(2)$ gauge field,
and the $F^{\prime}_{\mu\nu}={\bf F}^{\prime}_{\mu\nu}\cdot{\bf
T}=\partial_{\mu}({\bf A}_{\nu} \cdot {\bf T}
)-\partial_{\nu}({\bf A}_{\mu} \cdot {\bf T})-{ i} g[{\bf A}_{\mu}
\cdot {\bf T},{\bf A}_{\nu} \cdot {\bf T}]$ is its field strength.
Hence the field strength ${\bf F}^{\prime}_{\mu\nu}$ satisfies
\begin{equation}
\label{y0004} {\bf F}^{\prime}_{\mu\nu}=\partial_{\mu}{\bf
A}_{\nu}-\partial_{\nu}{\bf A}_{\mu}+g {\bf A}_{\mu}\times {\bf
A}_{\nu} ~.
\end{equation}
We use the $A^{0}_{\mu}$ denote the $U(1)$ gauge field. Its gauge
field strength can be expressed as follows
\begin{equation}
\label{y0005} E^{\prime}_{\mu\nu}=\partial_{\mu}
A^{0}_{\nu}-\partial_{\nu} A^{0}_{\mu}~.
\end{equation}

In quantum theory Pauli matrices are usually represented as the
following
\begin{eqnarray}
\label{y0006}\tau_{1}= \left(
\begin{array}{cc}
0 &  1
\\
1 & 0
\end{array}
\right)~,~~\tau_{2}= \left(
\begin{array}{cc}
0 &  -i
\\
i & 0
\end{array}
\right)~,~~\tau_{3}= \left(
\begin{array}{cc}
1 &  0
\\
0 & -1
\end{array}
\right)~.
\end{eqnarray}
The generators of $SU(2)$ Lie algebra can be then written as
\begin{equation}
\label{y0008} T^{i}=\frac{1}{2}\tau^{i}~.
\end{equation}
Furthermore the $T^{i}$ obey the commutation relations of the
$SU(2)$ Lie algebra
\begin{equation}
\label{y0010}
[T^{i},T^{j}]=i\sum^{3}_{k=1}\varepsilon_{ijk}T^{k}~.
\end{equation}
Here $\varepsilon_{ijk}$ is the totally antisymmetry tensor in
3-dimensions. The spherical components of ${\bf T}$ are
\begin{eqnarray}
\label{y0012} T_{+}&=&T_{1}+i T_{2}=\left(
\begin{array}{cc}
0 & 1
\\
0 & 0
\end{array}
\right)~,
\\
\label{y0016}T_{-}&=&T_{1}-i T_{2}=\left(
\begin{array}{cc}
0 & 0
\\
1 & 0
\end{array}
\right)~,
\\
\label{y0018} T_{3}&=&\frac{1}{2}\tau_{3}~=~\frac{1}{2}\left(
\begin{array}{cc}
1 & 0
\\
0 & -1
\end{array}
\right)~.
\end{eqnarray}
Therefore the scalar product ${\bf A}_{\mu}\cdot {\bf T}$ in the
isospace can be expressed in terms of the spherical components of
${\bf T}$, e.g. ${\bf A}_{\mu}\cdot {\bf
T}=\frac{1}{\sqrt{2}}(A^{+}_{\mu}T_{+}+A^{-}_{\mu}T_{-})+A^{3}_{\mu}T_{3}$,
where the $A^{+}_{\mu}$ and $A^{-}_{\mu}$ are defined by
\begin{eqnarray}
\label{Apm} A^{\pm}_{\mu}=\frac{1}{\sqrt{2}}(A^{1}_{\mu}\mp
iA^{2}_{\mu})~,
\end{eqnarray}

The complete Lagrangian density without mass terms are gauge
invariant and thus will lead to a renormalizable quantum field
theory. But the Lagrangian density of Eq.(\ref{y0001+}) can not
describe the real physical world. To make the vector bosons in
$SU(2)$ gauge field obtain the masses, we introduce the
transformation
\begin{equation}
\label{y0020}
A^{i}_{\mu}=B^{i}_{\mu}-C^{i}_{\mu}~,~~~~~~~~i=0,1,2,3~.
\end{equation}
Where $C^{i}_{\mu}$ are four special vectors since every
components of them are $4\times4$ matrices. Thus every components
of the vector fields $B^{i}_{\mu}$ must be $4\times4$ matrices
either. Inserting Eq.(\ref{y0020}) in Eq.(\ref{Apm}) directly
yields
\begin{eqnarray}
 \label{y0034}
B^{\pm}_{\mu}=\frac{1}{\sqrt{2}}(B^{1}_{\mu}\mp iB^{2}_{\mu})~,
~C^{\pm}_{\mu}&=&\frac{1}{\sqrt{2}}(C^{1}_{\mu}\mp iC^{2}_{\mu})~,
\end{eqnarray}
and
\begin{equation}
\label{y1034}
B^{+}_{\mu}T_{+}+B^{-}_{\mu}T_{-}={\sqrt{2}}(B^{1}_{\mu}T_{1}+
B^{2}_{\mu}T_{2})~.
\end{equation}
Here we assume that $C^{i}_{\mu}$ can be expressed as follows
\begin{eqnarray}\nonumber
C^{-}_{\mu}&=&\frac{1}{2\sqrt{2}g}m_{21}\gamma_{\mu}~,~~C^{0}_{\mu}=\frac{1}{8
g^{\prime} }(m_{11}+m_{22})\gamma_{\mu} ~,
\\
\label{y0024+}
C^{+}_{\mu}&=&\frac{1}{2\sqrt{2}g}m_{12}\gamma_{\mu}~,~~C^{3}_{\mu}=\frac{1}{4
g}(m_{11}-m_{22})\gamma_{\mu} ~.
\end{eqnarray}
Where $m_{11},~m_{22}$ are real parameters, and $m_{12},~m_{21}$
satisfy $m_{12}^{*}=m_{21}$(here $*$ denotes complex conjugate).

Inserting the equations (\ref{y0020}), (\ref{y0034}),
(\ref{y1034}) and (\ref{y0024+}) in Eq.(\ref{y0001+}), and using
the identity, e.g. $\gamma^{\mu}\gamma_{\mu}=4$, one can rewrite
the Lagrangian density of Eq.(\ref{y0001+}) as follows
\begin{eqnarray}
\nonumber {\cal
L}&=&i{\bar{\psi}}\gamma^{\mu}\partial_{\mu}\psi+\frac{g}{\sqrt{2}}{\bar
{\psi}}\gamma^{\mu}B^{+}_{\mu}T_{+}\psi \\
\nonumber &+&\frac{g}{\sqrt{2}}{\bar
{\psi}}\gamma^{\mu}B^{-}_{\mu}T_{-}\psi+g{\bar
{\psi}}\gamma^{\mu}B^{3}_{\mu}T_{3}\psi \\
\nonumber & + &
g^{\prime}{\bar{\psi}}\gamma^{\mu}B^{0}_{\mu}\psi-{\bar{\psi}}{\hat
{M}}\psi
\\
\label{totallagrangian} &-&\frac{1}{4}~{\bf
F}^{\prime}_{\mu\nu}\cdot{\bf
F}^{\prime\mu\nu}-~\frac{1}{4}~E^{\prime}_{\mu\nu}E^{\prime\mu\nu}~.
\end{eqnarray}
Where the operator
\begin{eqnarray}
\label{mass}\hat{M}=\left(
\begin{array}{cc}
m_{11} & m_{12}
\\
m_{21} & m_{22}
\end{array}
\right)
\end{eqnarray}
is obviously a Hermitian operator, namely,
${\hat{M}}^{\dag}={\hat{M}}$. The term $-{\bar{\psi}}{\hat
{M}}\psi$ in Eq.(\ref{totallagrangian}) can be looked as the mass
term of the fermion fields. Hence the operator $\hat{M}$ is the
fermion mass matrix.

In Yang-Mills theory the isospin doublet $\psi$ can be transformed
by
\begin{equation}
\label{y0026} \psi\to\tilde{\psi}=U\psi=\exp (i{\bf  a}(x)\cdot
{\bf T})\psi~.
\end{equation}
Under the transformation Eq.(\ref{y0026}) of the isospin doublet,
the $SU(2)$ gauge field ${\bf A}_{\mu}\cdot{\bf T}$ transforms as
follows
\begin{equation}
\label{y0028} {\bf A}^{\prime}_{\mu}\cdot{\bf T}=U{\bf
A}_{\mu}\cdot{\bf T}U^{-1}+\frac{i}{g}U(\partial_{\mu}U^{-1})~.
\end{equation}
After we do the transformation of Eq.(\ref{y0020}), the invariant
requirement of the lagrangian of Eq.(\ref{y0001+}) under the local
$SU(2)\times U(1)$ gauge transformations makes sure that
\begin{equation}
\label{y0030} {\bf C}^{\prime}_{\mu}\cdot{\bf T}=U{\bf
C}_{\mu}\cdot{\bf T}U^{-1}~.
\end{equation}
Inserting Eq.(\ref{y0020}) in Eq.(\ref{y0028}) and minus
Eq.(\ref{y0030}) then yields
\begin{equation}
\label{y0032} {\bf B}^{\prime}_{\mu}\cdot{\bf T}=U{\bf
B}_{\mu}\cdot{\bf T}U^{-1}+\frac{i}{g}U(\partial_{\mu}U^{-1})~.
\end{equation}
It is intriguing that the ${\bf B}_{\mu}$ is also an $SU(2)$
Yang-Mills field.

Following the same sequence as the above argument on the local
$SU(2)$ gauge transformations, one can similarly prove that
$B_{\mu}^{0}$ should be treated as a $U(1)$ gauge field provided
that the transformation matrix $U=\exp (i{\bf  a}(x)\cdot {\bf
T})$ is substituted by the phase factor $U=\exp (i\theta(x))$ in
Eq.(\ref{y0026}) and the $C_{\mu}^{0}$ transforms as
\begin{equation}
\label{c0} ({ C}^{0}_{\mu})^{\prime}=C_{\mu}^{0}
\end{equation}
under the local $U(1)$ rotation. Hence one can define the field
strength of $B_{\mu}^{0}$ field as follows
\begin{equation}
\label{y0105} E_{\mu\nu}=\partial_{\mu} B^{0}_{\nu}-\partial_{\nu}
B^{0}_{\mu}~.
\end{equation}
Inserting Eq.(\ref{y0020}) and Eq.(\ref{y0024+}) in
Eq.(\ref{y0005}) directly yields
$E^{\prime}_{\mu\nu}=\partial_{\mu} B^{0}_{\nu}-\partial_{\nu}
B^{0}_{\mu}$. That is to say,
\begin{equation}
\label{fieldstrength}E_{\mu\nu}=E^{\prime}_{\mu\nu}~.
\end{equation}
Through Eq.(\ref{totallagrangian}) and Eq.(\ref{fieldstrength}),
we can draw a conclusion that the $B_{\mu}^{0}$ is a massless
vector field.

In complete Lagrangian of Eq.(\ref{totallagrangian}), the energy
density of the $SU(2)$ gauge field is
\begin{eqnarray}\label{lagrangiansu2}
{\cal L}_{SU(2)}=  -\frac{1}{4}~{\bf F}^{\prime}_{\mu\nu}\cdot{\bf
F}^{\prime\mu\nu}~.
\end{eqnarray}
In the following we will consider the change of the ${\cal
L}_{SU(2)}$ under the transformation from ${\bf A_{\mu}}$ to ${\bf
B_{\mu}}$. According to the equations (\ref{y0004}),
(\ref{y0020}), (\ref{y0034}) and (\ref{y1034}), we rewrite the
gauge field energy density of Eq.(\ref{lagrangiansu2})
as\footnote{We ignore the factor $-\frac{1}{4}$ in
Eq.(\ref{y0099}) for simplicity.}
\begin{widetext}
\begin{eqnarray}
\nonumber &&{\bf F}^{\prime}_{\mu\nu}\cdot{\bf F}^{\prime\mu\nu}
 =\sum^{3}_{i=1}
[\partial_{\mu}B^{i}_{\nu}-\partial_{\nu}B^{i}_{\mu}
+g\varepsilon_{ikl}(B^{k}_{\mu}-C^{k}_{\mu})(B^{l}_{\nu}-C^{l}_{\nu})]\times
 \\
\nonumber &&   [\partial^{\mu}B^{i\nu}-\partial^{\nu}B^{i\mu}+g
\varepsilon_{imn}(B^{m\mu}-C^{m\mu})(B^{n\nu}-C^{n\nu})]
\\
\nonumber
&&=2[(\partial_{\mu}B^{-}_{\nu}-\partial_{\nu}B^{-}_{\mu})-i
g(B^{-}_{\mu}B^{3}_{\nu}-B^{-}_{\nu}B^{3}_{\mu})-i g V_{\mu\nu}
]\times
\\
\nonumber &&[(\partial^{\mu}B^{+\nu}-\partial^{\nu}B^{+\mu})+i g
(B^{+\mu}B^{3\nu}-B^{+\nu}B^{3\mu})+i g X^{\mu\nu}]
+\\
\nonumber
 &&[(\partial_{\mu}B^{3}_{\nu}-\partial_{\nu}B^{3}_{\mu})
 +i g (B^{-}_{\mu}B^{+}_{\nu}-B^{-}_{\nu}B^{+}_{\mu})+i gY_{\mu\nu}]\times
 \\
\nonumber
 &&[(\partial^{\mu}B^{3\nu}-\partial^{\nu}B^{3\mu})+i g
 (B^{-\mu}B^{+\nu}-B^{-\nu}B^{+\mu})+i g Z^{\mu\nu}]
 \\
 \nonumber
&&={\bf F}_{\mu\nu}\cdot{\bf
F}^{\mu\nu}+2ig(\partial_{\mu}B^{-}_{\nu}-\partial_{\nu}B^{-}_{\mu})X^{\mu\nu}+2g^2
(B^{-}_{\mu}B^{3}_{\nu}-B^{-}_{\nu}B^{3}_{\mu})X^{\mu\nu}
\\
\nonumber &&
-2ig(\partial^{\mu}B^{+\nu}-\partial^{\nu}B^{+\mu})V_{\mu\nu}
+2g^2(B^{+\mu}B^{3\nu}-B^{+\nu}B^{3\mu})V_{\mu\nu}+2g^2V_{\mu\nu}X^{\mu\nu}
\\
\nonumber && + i
g(\partial_{\mu}B^{3}_{\nu}-\partial_{\nu}B^{3}_{\mu})Z^{\mu\nu}
-g^2(B^{-}_{\mu}B^{+}_{\nu}-B^{-}_{\nu}B^{+}_{\mu})Z^{\mu\nu}-g^2Y_{\mu\nu}Z^{\mu\nu}
\\
\label{y0099} && +i g
(\partial^{\mu}B^{3\nu}-\partial^{\nu}B^{3\mu})Y_{\mu\nu}
-g^2(B^{-\mu}B^{+\nu}-B^{-\nu}B^{+\mu})Y_{\mu\nu} ~.
\end{eqnarray}
Where
\begin{equation}
\label{y0104} {\bf F}_{\mu\nu}=\partial_{\mu}{\bf
B}_{\nu}-\partial_{\nu}{\bf B}_{\mu}+g {\bf B}_{\mu}\times {\bf
B}_{\nu} ~.
\end{equation}
The Yang-Mills theory shows that the ${\bf F}_{\mu\nu}$ is the
field strength of the $SU(2)$ gauge field ${\bf B}_{\mu}$. The
tensors $V_{\mu\nu}, X^{\mu\nu}, Y_{\mu\nu}$ and $Z^{\mu\nu}$ are
all the bilinear functions of $ B^{i}_{\mu}$ and $C^{i}_{\mu}$
\begin{eqnarray}
\label{y0060} &&
V_{\mu\nu}=-C^{-}_{\mu}B^{3}_{\nu}-B^{-}_{\mu}C^{3}_{\nu}+C^{-}_{\mu}C^{3}_{\nu}
+C^{-}_{\nu}B^{3}_{\mu}+B^{-}_{\nu}C^{3}_{\mu}-C^{-}_{\nu}C^{3}_{\mu}~,
\\
\label{y0062} &&
X^{\mu\nu}=-C^{3\nu}B^{+\mu}-B^{3\nu}C^{+\mu}+C^{+\mu}C^{3\nu}
+C^{+\nu}B^{3\mu}+B^{+\nu}C^{3\mu}-C^{+\nu}C^{3\mu}~,
\\
\label{y0070} &&
Y_{\mu\nu}=-C^{-}_{\mu}B^{+}_{\nu}-B^{-}_{\mu}C^{+}_{\nu}+C^{-}_{\mu}C^{+}_{\nu}
+C^{-}_{\nu}B^{+}_{\mu}+B^{-}_{\nu}C^{+}_{\mu}-C^{-}_{\nu}C^{+}_{\mu}~,
\\
\label{y0072} &&
Z^{\mu\nu}=-C^{-\mu}B^{+\nu}-B^{-\mu}C^{+\nu}+C^{-\mu}C^{+\nu}
+C^{-\nu}B^{+\mu}+B^{-\nu}C^{+\mu}-C^{-\nu}C^{+\mu}~.
\end{eqnarray}
\end{widetext}
From Eq.(\ref{lagrangiansu2}) and Eq.(\ref{y0099}), one can obtain
the following mass terms in the Lagrangian for the $B^{3}_{\mu}$,
$B^{+}_{\mu}$ and $B^{-}_{\mu}$ fields
\begin{eqnarray}
\label{massterm1} {\cal
L}_{mass3}&=&-\frac{3}{8}m_{12}m_{21}B^{3}_{\mu}B^{3\mu}~,
\\
\label{massterm2} {\cal
L}_{mass+}&=&\frac{3}{16}m_{21}m_{21}B^{+}_{\mu}B^{+\mu}~,
\\
\label{massterm3} {\cal
L}_{mass-}&=&\frac{3}{16}m_{12}m_{12}B^{-}_{\mu}B^{-\mu}~.
\end{eqnarray}
The standard mass term in a Lagrangian for any massive vector
field $V_{\mu}(x)$ is given by ${\cal L}_{m}=\frac{1}{2}m^2
V_{\mu}V^{\mu}$. Therefore above equations demonstrate that the
$B^{3}_{\mu}$, $B^{+}_{\mu}$ and $B^{-}_{\mu}$ fields are all
massive provided that
\begin{eqnarray}
\label{bosonmass} m_{3}^{2}=-\frac{3}{4}m_{12}m_{21}~,~~
m_{+}^{2}=\frac{3}{8}m_{21}^{2}~,~~
m_{-}^{2}=\frac{3}{8}m_{12}^{2}~.
\end{eqnarray}
It is intriguing that the mass terms of the intermediate vector
bosons are unrelated with the diagonal elements of the fermion
mass matrix. One can obtain the following interesting formula from
Eq.(\ref{bosonmass})
\begin{eqnarray}
\label{bosonmass++} \frac{3}{8} (m_{12}-m_{21})^{2}=m_{3}^{2}+
m_{+}^{2}+ m_{-}^{2}~.
\end{eqnarray}
Thus Eq.(\ref{bosonmass}) and Eq.(\ref{bosonmass++}) reveal the
connection between the masses of gauge bosons and the fermion mass
matrix.

More generally, the fermion mass matrix must not satisfy the
Hermitian condition, that is, ${\hat{M}}^{\dag}\neq{\hat{M}}$. The
most general mass term consistent with fermion number conservation
has the form
\begin{eqnarray}
\label{generalmass} -{\bar{\psi}}_{L}{\hat
{M}}\psi_{R}-{\bar{\psi}}_{R}{\hat {M}}^{\dag}\psi_{L}~,
\end{eqnarray}
where ${\hat {M}}$ is a completely general complex matrix in the
isospin space, $\psi_{L}$ and $\psi_{R}$ are the ``left-handed"
and the ``right-handed" fields and defined by
\begin{eqnarray}\label{leftright}
\psi_{L}=P_{-}\psi~,~~~~\psi_{R}=P_{+}\psi~,
\end{eqnarray}
where $P_{\pm}=\frac{1}{2}(1\pm\gamma_{5})$ are the projection
operators. In this general case, we can select an interesting
ansatz, e.g. $m_{12}=-m_{21}=m$, then we acquire
\begin{eqnarray}\label{massend}
m_{3}=\frac{\sqrt{3}}{2}m~,~~
m_{+}=\frac{\sqrt{3}}{2\sqrt{2}}m~,~~
m_{-}=\frac{\sqrt{3}}{2\sqrt{2}}m~.
\end{eqnarray}

The equation (\ref{y0099}) obviously demonstrates that there are
constant terms in the lagrangian of Eq.(\ref{lagrangiansu2}),
which can be expressed in terms of $C_{\mu}^{i}$
\begin{eqnarray}\nonumber
&&{\cal L}_{const}=
2g^{2}(C^{-}_{\mu}C^{3}_{\nu}-C^{-}_{\nu}C^{3}_{\mu})(C^{+\mu}C^{3\nu}-C^{+\nu}C^{3\mu})
\\
\label{const}
&&-g^{2}(C^{-}_{\mu}C^{+}_{\nu}-C^{-}_{\nu}C^{+}_{\mu})(C^{-\mu}C^{+\nu}-C^{-\nu}C^{+\mu})~.
\end{eqnarray}
One can easily find that ${\cal L}_{const}=0$. Hence the special
fields $C^{k}_{\mu}(k=+,-,3)$ don't affect the energy density of
vacuum.

Since the gauge field determined  by the Lagrangian of
Eq.(\ref{y0001+}) satisfies the gauge invariance and involves
massless vector bosons $A^{i}_{\mu}$, it is renormalizable. In our
mechanism, it is proved that the gauge invariance of $SU(2)\times
U(1)$ is kept in the transformation of Eq.(\ref{y0020}); the
fermion mass term appears naturally; the masses of intermediate
vector bosons originate from the coupling between the gauge boson
fields and the fields $C^{i}_{\mu}(x)$. Similar to the standard
model with the Higgs mechanism, the gauge theory proposed in this
letter should be renormalizable either.

As shown by Fadeev and Popov \cite{fp67}, Feynman graphs including
at least one loop diagram of the gauge boson field are no longer
gauge invariant in Yang-Mills theory. The reason for this failure
is the fact that the Feynman rules describing Yang-Mills theory as
obtained so far are incomplete. One must take into account the
additional contributions of the so-called ghost fields. These
fields are described by the additional Lagrangian
\begin{eqnarray}
\label{fadeev} {\cal L}_{ghost}=-{\bar{\chi}}^{k}{\hat {\bf
L}}^{\mu}(\partial_{\mu}\delta_{kl}+g\varepsilon_{klm}B^{m}_{\mu})\chi^{l}~,
\end{eqnarray}
the ghost fields are denoted by $\chi^{k}$, where ${\hat {\bf
L}}^{\mu}$ is an arbitrary linear operator, which appears in the
gauge condition ${\hat {\bf L}}^{\mu}B^{k}_{\mu}=0$. Taking the
Lorentz gauge $\partial^{\mu}B^{k}_{\mu}=0$, one get
\begin{eqnarray}
\label{fadeev++} {\cal
L}_{ghost}=-{\bar{\chi}}^{k}\square\chi^{k}
-g\varepsilon_{klm}B^{m}_{\mu}{\bar{\chi}}^{k}
\partial_{\mu}\chi^{l}~.
\end{eqnarray}
These demonstrate that there are ghost-boson couplings in
Yang-Mills theory. In the lagrangian of Eq.(\ref{y0099}), it is
easy to find the couplings between the massive gauge boson fields
and the fields $C^{k}_{\mu}(x)$. Moreover, ghost propagators may
not occur as external lines of a Feynman diagram, since they do
not correspond to the physical modes. In this aspect the fields
$C^{k}_{\mu}(x)$ are the same as the ghost fields. More concise
discussion on the relation between the ghost fields and
$C^{k}_{\mu}(x)$ will be given in forthcoming paper.

For explaining the idea explicitly, we construct the theory under
the primary framework of Yang-Mills field. The parity violation
and the down quark mixing can not be included in this frame. The
question why nature select the massive fermions and the massive
SU(2) gauge bosons can not be answered yet.

In this letter, a new mechanism generating the massive
intermediate vector bosons in Yang-Mills theory is proposed
without introducing the Higgs boson. The relation between the
masses of the gauge bosons and the fermion mass matrix is
revealed. At the last of this letter, the properties of
$C^{k}_{\mu}(x)$ are discussed.

\begin{acknowledgments}
I am grateful to Prof. B. Q. Ma, Prof. J. Ng and B. Wu for their
enlightening discussion.

\end{acknowledgments}

\end{document}